\documentclass[aps,pra,twocolumn,groupedaddress,showpacs]{revtex4}
\usepackage{}
\usepackage{graphicx}
\usepackage{mathrsfs}

\begin{document}

\title{Optimal Dynamical Decoupling Sequence for Ohmic Spectrum}

\author{Yu Pan$^{1,2}$}
\author{Zai-Rong Xi$^1$}
\email{zrxi@iss.ac.cn}
\author{Wei Cui$^{1,2}$}

\affiliation{$^1$Key Laboratory of Systems and Control, Institute of
Systems Science, Academy of Mathematics and Systems Science, Chinese
Academy of Sciences, Beijing 100190, People's Republic of China}
\affiliation{$^2$Graduate University  of Chinese Academy of
Sciences, Beijing 100039, People's Republic of China}

\begin{abstract}
  We investigate the optimal dynamical decoupling sequence for a qubit coupled to an ohmic
  environment. By analytically computing the derivatives of the decoherence function, the optimal pulse
   locations are found to satisfy a set of non-linear equations which can be easily
   solved. These equations incorporates the environment information
   such as high-energy (UV) cutoff frequency $\omega_c$, giving a complete
   description of the decoupling process. The solutions explain previous experimental and theoretical
   results of locally optimized dynamical decoupling (LODD) sequence in high-frequency-dominated environment,
which were obtained by purely numerical computation and
    experimental feedback. As shown in numerical comparison, these solutions outperform the Uhrig dynamical decoupling
    (UDD)
    sequence by one or more orders of magnitude in the ohmic case.
\end{abstract}


\maketitle

\section{INTRODUCTION}
  Suppressing decoherence is one of the fundamental issues in the field of quantum information
  processing. Decoherence, which has been caused by the
  environmental noise, plagues almost all the implementations of quantum
  bit. To eliminate the unwanted coupling between a qubit and its
  environment, several schemes have been proposed and tested. Among them a promising one is
  dynamical decoupling \cite{ar1,ar2,ar3,ar4}, which restores the qubit coherence by
  applying delicately designed sequence of control pulses.\par
  For a qubit that can be modeled by a spin-1/2 particle, the
oldest dynamical decoupling sequence is periodic dynamical
decoupling (PDD). Originated from pulse sequences widely used in
nuclear magnetic resonance (NMR) \cite{ar5}, the PDD sequence
consists of periodic and equidistant $\pi$ pulses. To
  achieve better performance, there has
  been an extensive study in how to optimize the pulse locations \cite{ar6,ar7,ar8,ar9,ar10,ar11,ar12,ar13,ar14,ar15}. One important progress is the powerful Uhrig DD
  (UDD) \cite{ar10}, which employs $n$ pulses located at $t_j$ according to the simple rules$$\delta_j=\sin^2(j\pi/2(n+1)),$$where $\delta_j=t_j/T$ and $T$ is the total evolution time. UDD is
  first derived on spin-boson model and further proved to be universal in the sense that it can remove the qubit-bath coupling to $n^{th}$
  order in generic environment \cite{ar11}.\par
  Beyond UDD, another locally optimized dynamical decoupling (LODD)
  sequence has drawn great attention \cite{ar14}. LODD, along with its simplified
  version optimized noise-filtration dynamic decoupling
  (OFDD) \cite{ar15}, generates the decoupling
sequence by directly optimizing the
  decoherence function using numeric methods as well as experimental feedback. It has been shown to be
  able to suppress decoherence effect by orders of magnitude over
  UDD for certain noise spectrum, especially for the one with a high frequency part and
  sharp high-energy (UV)
  cutoff.\par
  However, in spite of the great experimental success, analytical results about the LODD sequence is insufficient. Until recently S. Pasini and G. S. Uhrig has made an analytical
  progress in optimizing the decoherence function for power law
  spectrum (PLODD) \cite{ar13}. The power law spectrum $\omega^{\alpha}$ for $\alpha<1$ without UV cutoff is
  considered. They minimize the decoherence function through expanding the function and separating, canceling divergences from the relevant terms and solving
variation problems. Inspired by Pasini's work, we try to analyze the
LODD problem
  with respect to the ohmic spectrum $S(\omega)\sim\omega$ and a sharp UV cutoff. Ohmic noise is the major decoherence source often found in a qubit's
   environment, for example, the semiconducting quantum dot \cite{ar16} and superconducting qubit \cite{ar17}. Optimal performance pulse sequence is found analytically which
   entirely
  differs from the UDD sequence in such environment. We call this kind of optimal sequence HLODD (LODD for ohmic spectrum) for short.\par
  We organize this paper as follows. In the second section we propose the optimization problem of
  the decoherence function. In section III, we derive the
  analytical equations for the optimal pulse sequence. In the
  following section, we run a simulation to verify our results.
  Conclusions are put in section V.

\section{OPTIMIZATION OF THE DECOHERENCE FUNCTION}
  Given a two-level quantum system, when the environmental noise behaves quantum-mechanically, we use the long-established spin-boson model with pure
  dephasing$$H=\sum_i\omega_ib_i^{\dagger}b_i+\frac{1}{2}\sigma_{z}\sum_i\lambda_i(b_i^{\dagger}+b_i). \eqno(1)$$Here
  we ignore the qubit free evolution hamiltonian. On the other hand, when the qubit is subjected to classical
  noise, the system is modeled
  as \cite{ar18,ar19}$$H=\frac{1}{2}[\Omega+\beta(t)]\sigma_z, \eqno(2)$$where the $\Omega$ is
  the qubit energy splitting and $\beta(t)$ the classical random noise. Let $t$ be the total evolution time, and $n$ pulses
   are applied at $t_1<t_2<...<t_n$ in sequence with negligible pulse durations . We use the notation $\delta_j=\frac{t_j}{t}$. This naturally leads
    to the definition of $t_0=0$ and $t_{n+1}=1$. In
  either ($1$) or ($2$), the decay of coherence under the dynamical decoupling sequence can be described by the
  decoherence function \cite{ar10,ar18,ar20,ar21} $e^{-2\chi(t)}$
  with$$\chi(t)=\int_0^{\infty}\frac{S(\omega)}{\omega^2}{|y_n(\omega{t})|}^2d\omega, \eqno(3)$$where
  $S(\omega)$ is environmental noise spectrum. The filter function
  $y_n(t)$ is given
  by$$y_n(t)=1+(-1)^{n+1}e^{\mbox{i}\omega{t}}+2\sum_{j=1}^n(-1)^je^{\mbox{i}\omega{t}\delta_j}. \eqno(4)$$Thus minimization
  of $\chi(t)$ with respect to $\delta_j$ gives the optimal
  decoupling sequence.\par
  We now consider the case when the noise spectrum is ohmic with a sharp cutoff at $\omega_c$, i.e.
  $S(\omega)=S_0\omega\Theta(\omega_c-\omega)$. $S_0$ is an irrelevant constant factor and $\Theta$ is unit step function.
  Then minimization of ($3$) turns to minimization of
  $I_n$ with$$I_n=\int_0^{z_c}\frac{{|y_n(z)|}^2}{z}dz, \eqno(5)$$where $z_c=\omega_c{t}$. Since
  $y_n(0)=0$, the IR convergence insures the integral converges to a finite value \cite{ar13}.

\section{DERIVATION OF OPTIMAL PULSE SEQUENCE}
We follow the approach of Pasini and Uhrig \cite{ar13} to treat the
integral ($5$). Here we use notation
\[
  q_j=\left\{
     \begin{array}{ll}
        0 &\mbox{if $j = 0,n+1$,}\\
        1 &\mbox{if $j \in \{1,2,...,n\}$,}
     \end{array}
     \right.
\]
and$$\Delta_{ij}=\mbox{i}(\delta_i-\delta_j),$$from which we
get$$|y_n(z)|^2=\sum_{i,j=0}^{n+1}2^{q_i+q_j}(-1)^{i+j}e^{z\Delta_{ij}}.$$Then
the integral $I_n$ can be expressed as
$$I_n=\lim_{x\to0^+}I_n(x),$$$$I_n(x)=\sum_{i,j=0}^{n+1}2^{q_i+q_j}(-1)^{i+j}I_{ij}(x), \eqno(6)$$where
 the integrals $I_{ij}(x)$ are
\setcounter{equation}{6}
\begin{eqnarray}
 I_{ij}(x)&=&\int_x^{z_c}\frac{e^{\Delta_{ij}z}}{z}dz\nonumber\\
          &=&\int_{-\Delta_{ij}x}^{-\Delta_{ij}z_c}\frac{e^{-z}}{z}dz.
\end{eqnarray}
The limit $x\to0^+$ is carried out because $I_{ij}(0)$ does not
exist for arbitrary $i,j$. Making use of the series representation
of exponential function \cite{ar22}
\begin{eqnarray}
E_1(z)&=&\int_z^{\infty}\frac{e^{-t}}{t}dt\nonumber\\
&=&-\gamma-\ln{z}+\sum\limits_{k=1}^{\infty}\frac{(-1)^{k+1}}{k!k}z^k,\nonumber
\end{eqnarray}
where $\gamma$ is the Euler-Mascheroni constant and the sum
converges for all the complex $z$, $I_{ij}(x)$ can be written as
\setcounter{equation}{7}
\begin{eqnarray}
I_{ij}(x)&=&E_1(-\Delta_{ij}x)-E_1(-\Delta_{ij}z_c)\nonumber\\
&=&\ln{{(z_c}/{x})}+\sum\limits_{k=1}^{\infty}\frac{\Delta_{ij}^k}{k!k}(z_c^k-x^k).
\end{eqnarray}
Since we always have $y_n(0)=0$ which
implies$$|y_n(0)|^2=\sum_{i,j=0}^{n+1}2^{q_i+q_j}(-1)^{i+j}=0,
\eqno(9)$$we can now proceed by taking the limit $x\to0+$ in $I_n$
\setcounter{equation}{9}
\begin{eqnarray}
I_n&=&\lim_{x\to0^+}\sum_{i,j=0}^{n+1}2^{q_i+q_j}(-1)^{i+j}[\ln{{(z_c}/{x})}+\sum\limits_{k=1}^{\infty}\frac{\Delta_{ij}^k}{k!k}(z_c^k-x^k)]\nonumber\\
&=&\lim_{x\to0^+}\sum_{i,j=0}^{n+1}2^{q_i+q_j}(-1)^{i+j}\sum\limits_{k=1}^{\infty}\frac{\Delta_{ij}^k}{k!k}(z_c^k-x^k)\nonumber\\
&=&\sum_{i,j=0}^{n+1}\sum\limits_{k=1}^{\infty}2^{q_i+q_j}(-1)^{i+j}\frac{\Delta_{ij}^k}{k!k}z_c^k.
\end{eqnarray}
To minimize $I_n$, UDD requires the first $n$ derivatives of $y_n$
vanish while OFDD simplifies the optimization process by replacing
$S(\omega)$ by a constant. Here we attempt to minimize $I_n$
directly to obtain optimal pulse sequence. We notice that at the
optimal pulse locations $\delta_j \ (j=1,2,...,n)$, the gradient of
$I_n$ vanishes. So we impose the following conditions
$\frac{\partial{I_n}}{\partial\delta_m}=0,$ for $m$ from $1$ to $n$.
Although ($10$) are complex infinite series, we can still explicitly
compute the derivatives of ($10$) as long as these derivatives
converge. For arbitrary $m$ we have \setcounter{equation}{10}
\begin{eqnarray}
\frac{\partial{I_n}}{\partial\delta_m}&=&\frac{\partial}{\partial\delta_m}\sum_{i,j=0}^{n+1}\sum\limits_{k=1}^{\infty}2^{q_i+q_j}(-1)^{i+j}\frac{\Delta_{ij}^k}{k!k}z_c^k\nonumber\\
&=&\frac{\partial}{\partial\delta_m}\{\sum_{i=0}^{n+1}\sum\limits_{k=1}^{\infty}2^{q_i+q_m}(-1)^{i+m}\frac{\Delta_{im}^k}{k!k}z_c^k\nonumber\\
&&+\sum_{i=0}^{n+1}\sum\limits_{k=1}^{\infty}2^{q_m+q_i}(-1)^{m+i}\frac{\Delta_{mi}^k}{k!k}z_c^k\}\nonumber\\
&=&\sum_{i=0}^{n+1}\sum\limits_{k=1}^{\infty}2^{q_m+q_i}(-1)^{m+i}\frac{z_c^k}{k!}\mbox{i}[\Delta_{mi}^{k-1}-(-1)^{k-1}\Delta_{mi}^{k-1}].\nonumber\\
&&
\end{eqnarray}
The terms with $k$ odd cancel, so the result can be simplified as
\setcounter{equation}{11}
\begin{eqnarray}
&=&\sum_{i=0}^{n+1}\sum\limits_{k=1}^{\infty}2^{q_m+q_i+1}(-1)^{m+i}\frac{z_c^{2k}}{(2k)!}{\mbox{i}}^{2k}{(\delta_m-\delta_i)}^{2k-1}\nonumber\\
&=&\sum_{i\ne{m}}^{n+1}\frac{1}{\delta_m-\delta_i}2^{q_m+q_i+1}(-1)^{m+i}\sum\limits_{k=1}^{\infty}\frac{z_c^{2k}}{(2k)!}{\mbox{i}}^{2k}{(\delta_m-\delta_i)}^{2k}\nonumber\\
&=&\sum_{i\ne{m}}^{n+1}\frac{1}{\delta_m-\delta_i}2^{q_m+q_i+1}(-1)^{m+i}\{\cos[(\delta_m-\delta_i)z_c]-1\}.\nonumber\\
&&
\end{eqnarray}
Here we have used the expansion
$\cos{(z)}=\sum\limits_{k=0}^{\infty}\frac{(-1)^k}{(2k)!}z^{2k}$
which converges on the whole complex plane. From ($12$) we know that
the derivatives of $I_n$ indeed converge to a finite value. Thus the
optimal pulse locations $\{\delta_1,\delta_2,...\delta_n\}$ shall
satisfy the following non-linear equations
$$\sum_{i\ne{m}}^{n+1}\frac{1}{\delta_m-\delta_i}2^{q_m+q_i+1}(-1)^{m+i}\{\cos[(\delta_m-\delta_i)z_c]-1\}=0.
\eqno(13)$$Equations ($13$) are main results of this paper. The
optimal sequence obtained from ($13$) is quite different from the
UDD sequence obeying
$$\sum\limits_{j=1}^{n+1}2^{q_j}(-1)^j\delta_j^p=0$$ for
$p=\{1,2,...n\}$. For the ohmic spectrum, our equations incorporate
the UV cutoff frequency $\omega_c$, indicating that the solutions
are specially tailored to combat this kind of noise. Although the
UDD sequence is universal in suppressing decoherence, we believe
that the HLODD sequence will outperform the UDD sequence in the
ohmic environment. In the next section, we use numeric methods to
illustrate the performance of HLODD sequence.

\begin{figure}
\scalebox{0.5}[0.5]{\includegraphics{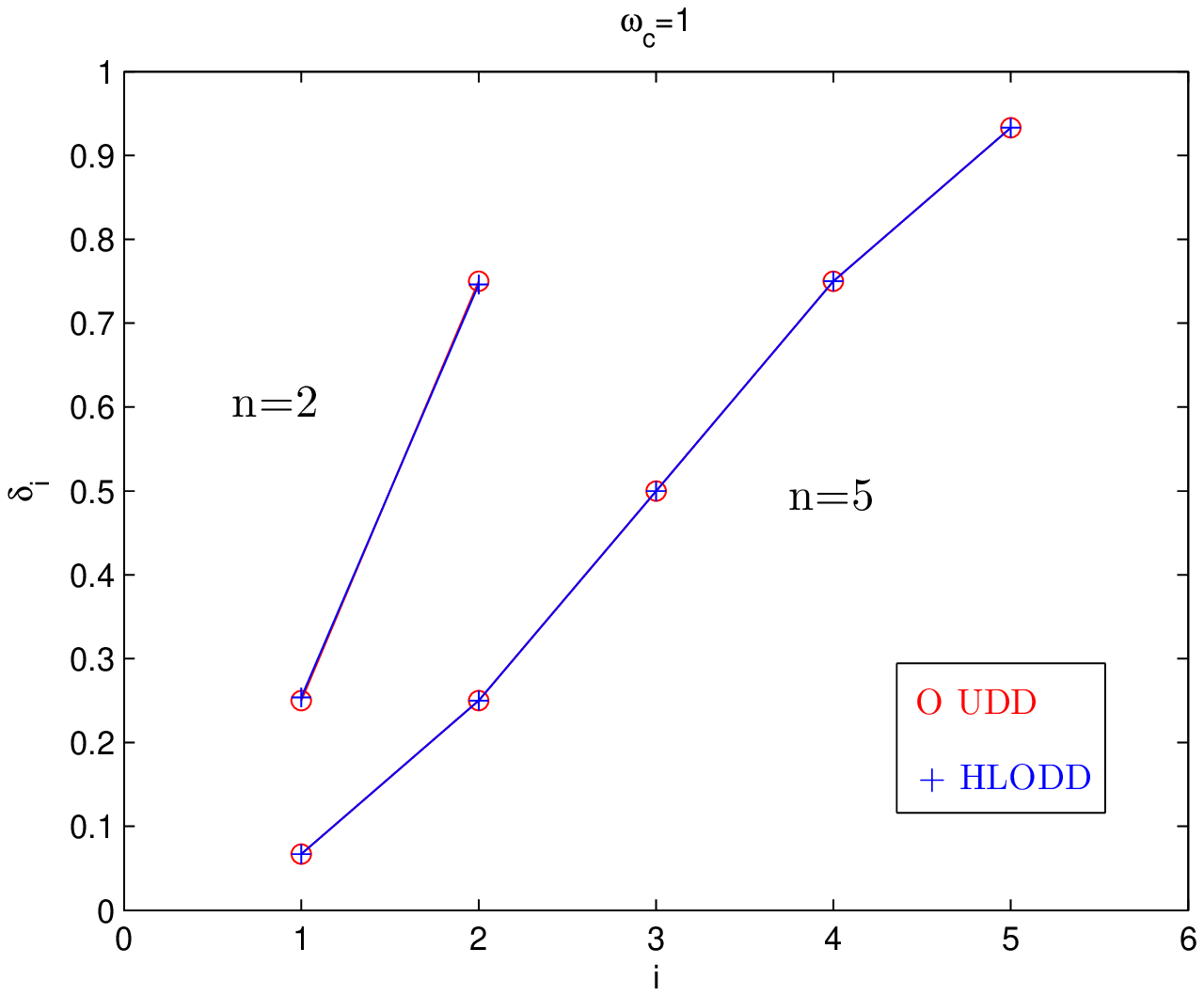}}
\scalebox{0.5}[0.5]{\includegraphics{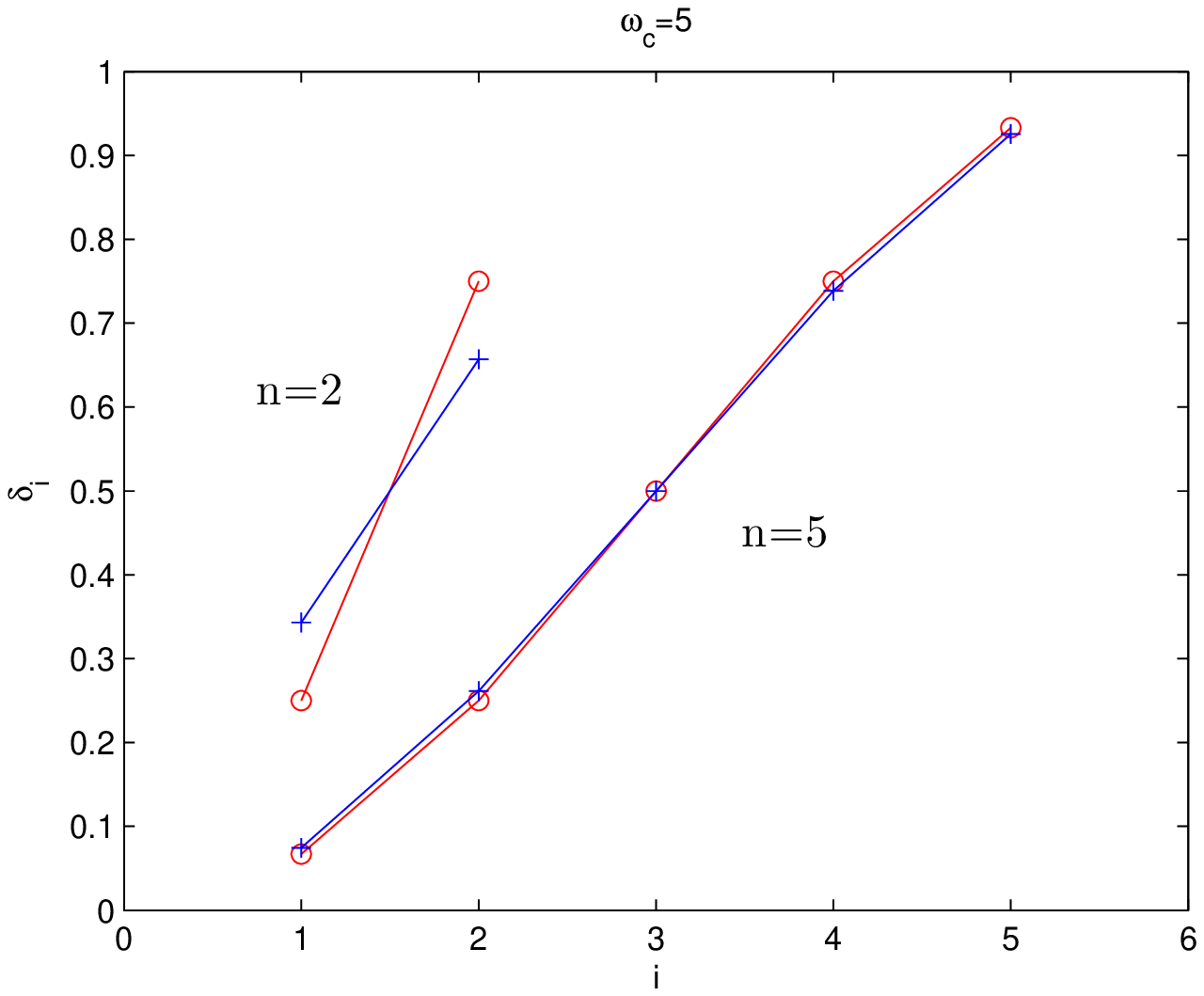}}
\scalebox{0.5}[0.5]{\includegraphics{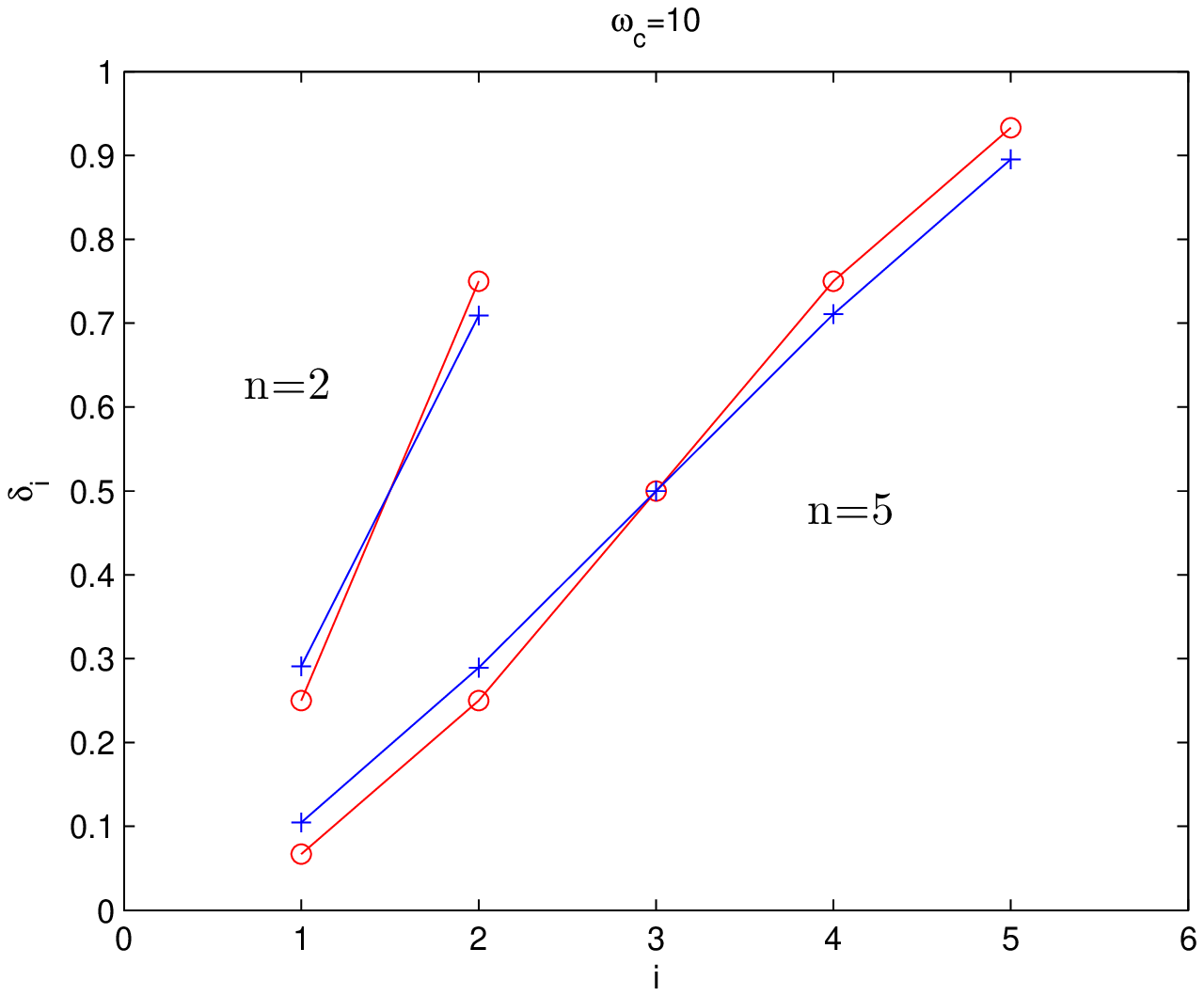}}\caption{Comparison
between UDD and HLODD sequence for different UV cutoff frequency
$\omega_c$. Pulse sequences $\delta_i$ for $n=2$ and $n=5$ are
plotted in one figure under the same $\omega_c$.\label{fig1}}
\end{figure}

\begin{figure}
\scalebox{0.6}[0.6]{\includegraphics{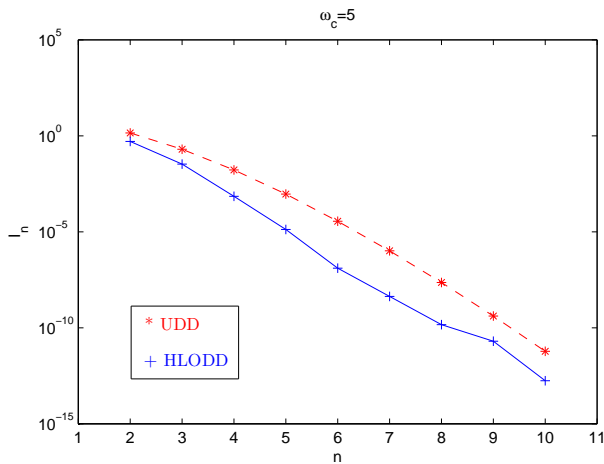}}\caption{Semi-log
plot of $I_n$ versus pulse number $n$. UDD and HLODD sequences are
compared.{\label{fig2}}}
\end{figure}

\section{NUMERICAL RESULTS}

  We start our simulation by solving the non-linear equations ($13$) and evaluate
   the decoherence function with these solutions. First, we set the total evolution time
  $t=1$ without loss of generality. Then $z_c=\omega_c$ and we can concentrate on analyzing the influence
  of the cutoff frequency $\omega_c$. Computing solutions to ($13$) for different $\omega_c$, we find that the
  optimal pulse sequences behave differently. We also evaluate the
  UDD sequence for comparison.\par
  As shown in Fig. \ref{fig1}, deviation  of the
  pulse locations $\delta_i$ in HLODD sequence from their UDD counterparts
  increases with $\omega_c$. This agrees with our intuition
  since UDD focuses on suppressing decoherence by minimizing $|y_n(z)|$ in the neighborhood of  $y_n(0)$, weakening
   its ability to maintain small $|y_n(z)|$ on the other end of the
   spectrum. For large $\omega_c$, UDD sequence is no longer
   optimal. In addition, we can see pulse number $n$ plays an important role.
   By increasing $n$, UDD can narrow the difference from HLODD. The difference between the two sequences when
   $n=2$ is greatly reduced when we increase $n$ to $5$, see Fig.
   \ref{fig1}. Especially for the case $\omega_c=1$, the difference is completely removed. However, for
    larger $\omega_c$ this gap can't be removed by increasing $n$.\par
  Next, to demonstrate the optimal decoupling ability of HLODD sequence, we compute
  $I_n$ versus $n$ while $\omega_c$ is chosen to be $5$. The results are depicted in Fig. \ref{fig2}, and again
  are compared with UDD. The obtained solutions yield a significant improvement over UDD.
  For fixed $n$, the HLODD suppresses decoherence better than UDD by
  one or two orders of magnitude which is in agreement with the results in \cite{ar14,ar15}, where LODD and OFDD sequences are tested for $^9$Be$^+$ qubits in
   a penning ion trap and various spectrum. The qubit error rates are below $10^{-5}$ when $n>5$, and we see that HLODD is capable of  suppressing the error rates far below
    the Fault-Tolerance error threshold \cite{ar23} by increasing $n$. Furthermore, by inspecting the points on the HLODD curve, we expect the HLODD
  sequence suppresses decoherence in power law $n$ as UDD.\par
  At last, we would like to explain the numerical results in another
  way. If we fixed UV cutoff frequency $\omega_c$ at the beginning,
  and compare the HLODD performance for $t=1$, $t=5$, and $t=10$,
  the numerical results would be the same since $z_c$ did not
  change. So we can also conclude that for the same number of
  pulses $n$, HLODD will beat UDD with increasing total evolution
  time $t$.

\section{CONCLUSIONS}
  In this paper we analytically find the optimal pulse locations to decouple a qubit in an ohmic environment. By deriving the
  analytical expressions for the derivatives of decoherence
  function, we obtain a set of non-linear equations which the optimal
  pulse sequence must obey. These equations are completely different
  from UDD and are more accurate, because they incorporate the effect of UV cutoff frequency
  $\omega_c$.\par
  In our numerical simulation, the analytical results provide an improvement over UDD sequence by an order or two of magnitude, which
  is consistent with previous results in LODD and OFDD obtained by purely
  numerical minimization and experimental feedback. We have to mention that the pulse performance is influenced by the sharp UV cutoff
   frequency $\omega_c$ greatly. The larger the UV cutoff $\omega_c$, the more HLODD deviates from UDD. Early work \cite{ar13,ar14,ar18,ar20} has pointed
    out that for soft large UV cutoff, UDD performs even worse and LODD is still a better choice. However, the integral ($3$) for
    $S(\omega)$
     with a soft cutoff is hard to analyze.\par
  In conclusion, our work provides an analytical solution to optimal dynamical decoupling for ohmic case. Our derivation is based on ohmic spectrum, but we believe it can be
extended to super-ohmic case $S(\omega)\sim\omega^\alpha (\alpha>1)$
via slight modification.

\begin{acknowledgments}
This work was supported by the National Natural Science Foundation
of China (No. 60774099, No. 60821091), the Chinese Academy of
Sciences (KJCX3-SYW-S01), and by the CAS Special Grant for
Postgraduate Research, Innovation and Practice.
\end{acknowledgments}

\bibliography{ref}

\end{document}